\documentclass[11pt]{article}
\usepackage{moriond,epsfig}

\def\be{\begin{equation}}
\def\ee{\end{equation}}
\def\bea{\begin{eqnarray}}
\def\eea{\end{eqnarray}}

\begin{document}
\vspace*{4cm}
\title{Measurement of high-p$_\mathrm{T}$ hadrons at RHIC-PHENIX}

\author{T.~ISOBE for the PHENIX Collaboration}

\address{Graduate School of Science, University of Tokyo, 7-3-1 Hongo,
  Bunkyo-ku, Tokyo 113-0033, Japan}

\maketitle\abstracts{Measurements from the RHIC experiments show a strong
suppression in the yield of high-p$_\mathrm{T}$ single hadrons and a
clear reduction in strength of the di-jet signal in two-hadron
azimuthal-angle correlation functions in central Au+Au collisions at
RHIC.  
These measurements should provide direct information on the properties
of the medium within which hard-scattered partons propagate. 
The PHENIX preliminary results on single high-p$_\mathrm{T}$ hadron
production and on particle azimuthal correlations are shown.
The properties of the medium created in relativistic heavy ion
collisions are discussed based on the results.} 

\section{Introduction}

The PHENIX experiment~\cite{phenix} has been carried out at the
Relativistic Heavy Ion Collider~(RHIC) at Brookhaven National
Laboratory to find evidence of a phase transition from normal
nuclear matter to a Quark-Gluon Plasma~(QGP), a new phase of matter
consisting of de-confined quarks and gluons, and is expected to be
formed at very high energy densities above $\epsilon_c \approx$
1~GeV/fm$^3$.  

One of the most exciting results to date at RHIC is that the yield of
$\pi^0$ at high transverse momentum~($p_\mathrm{T}$) in central
$\sqrt{s_\mathrm{NN}}$ = 200~GeV Au+Au collisions is suppressed compared 
to the yield in p+p collisions scaled by $N_{coll}$, the number of
underlying nucleon-nucleon collisions~\cite{130gev,bib1}.
The suppression is a final state effect since it is absent in d+Au 
collisions~\cite{bib2}. 
The direct photon yield is not suppressed in Au+Au
collisions~\cite{dphoton}, meaning that the initial hard scattering
yield in Au+Au is well reproduced as the $N_{coll}$ scaled yield in p+p.
The observed hadron suppression is interpreted as a consequence of the
energy loss of hard scattered partons traversing the hot and dense
matter produced in central Au+Au collisions.    
This effect, known as jet quenching, is a possible signature for the
creation of QGP~\cite{wang}. 

\section{Measurement of high-p$_\mathrm{T}$ hadrons}

Identified hadrons that can be measured in PHENIX up to the highest
p$_\mathrm{T}$ are $\pi^0$ and $\eta$ via their 2$\gamma$ decay mode.  

\subsection{Nuclear modification factor}
The amount of nuclear modification in dense matter can be quantified by
the nuclear modification factor~(R$_\mathrm{AA}$), which is the ratio of
the measured yield to the expected yield from the scaled p+p yields, and
is defined as follows: 
\begin{eqnarray}
 R_\mathrm{AA}(p_\mathrm{T}) =
  \frac{d^2N_\mathrm{AA}/dp_\mathrm{T}d\eta}{T_\mathrm{AA}(b)d^2\sigma_\mathrm{NN}/dp_\mathrm{T}d\eta},
\end{eqnarray}
\noindent
where the numerator is the invariant $\pi^0$ yield in unit rapidity and
the denominator is the expected yield in p+p collision scaled by the
number of underlying nucleon-nucleon collisions, which in turn is
defined with the nuclear thickness function $T_\mathrm{AA}(b)$
multiplied by the total cross section~$\sigma_\mathrm{NN}$ with the
impact parameter $b$ representing collision centrality. 
If a hard-scattered parton goes through the bulk matter without any
nuclear effects, the R$_\mathrm{AA}$ is unity.

\begin{figure}[htb]
 \begin{center}
  \includegraphics[height=.38\textheight]{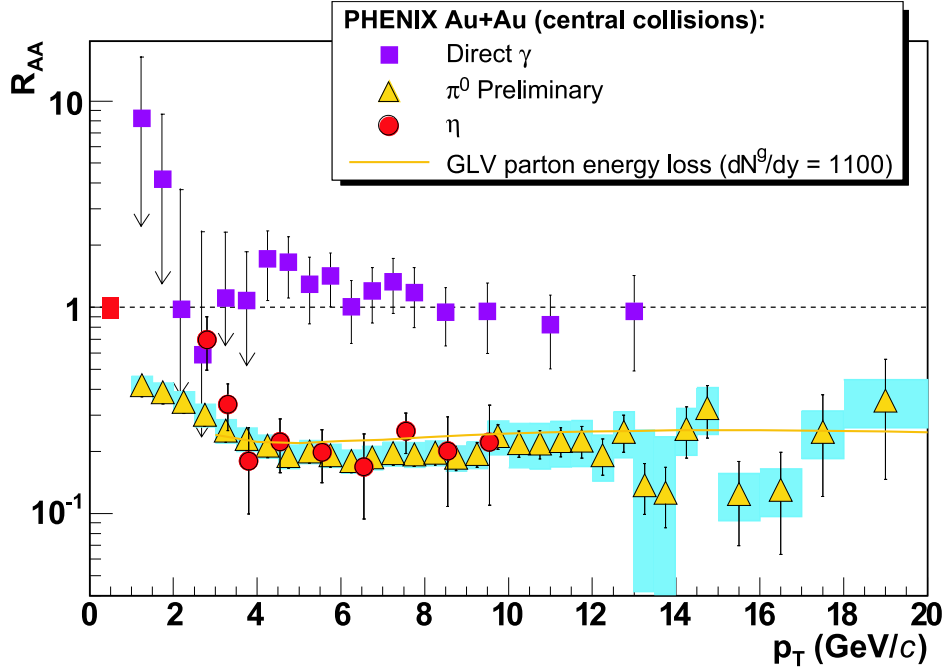}
  \caption{Nuclear modification factor, R$_\mathrm{AA}$ of $\pi^0$
 (triangles), $\eta$ (circles), and direct photon (squares). In addition
 to the statistical and p$_\mathrm{T}$-uncorrelated errors,
 point-to-point varying
 systematic errors are shown on the data points as boxes. An overall
 systematic error of $T_\mathrm{AA}$ normalization is shown at 1.}
 \label{fig:raa}
 \end{center}
\end{figure}

Figure~\ref{fig:raa} shows the preliminary Run-4 data of $\pi^0$
R$_\mathrm{AA}$ as a function of p$_\mathrm{T}$ along with final Run-2
results on $\eta$ and direct photon R$_\mathrm{AA}$~\cite{eta} in Au+Au
collisions at $\sqrt{s_\mathrm{NN}}$ = 200~GeV, as well as a theoretical
prediction which employs the GLV model~\cite{200glv}. 
As a result, a strong $\pi^0$ suppression by a factor of $\sim$ 5 is
observed, this suppression stays almost constant up to 20~GeV/$c$. 
The suppression pattern of $\eta$ is similar to that of $\pi^0$, and
this fact supports that the suppression occurs at partonic level.
The GLV model describes the strong suppression well and it indicates
the existence of bulk matter where the initial gluon density~(dN$^g$/dy)
is more than 1100, which corresponds to an energy density of
approximately 15~GeV/fm$^3$ in Au+Au collisions at
$\sqrt{s_\mathrm{NN}}$ = 200~GeV.   

\subsection{System-size dependence of jet quenching effect}

\begin{figure}[htb]
  \begin{minipage}{0.45\linewidth}
   \begin{center}
   \includegraphics[height=.3\textheight]{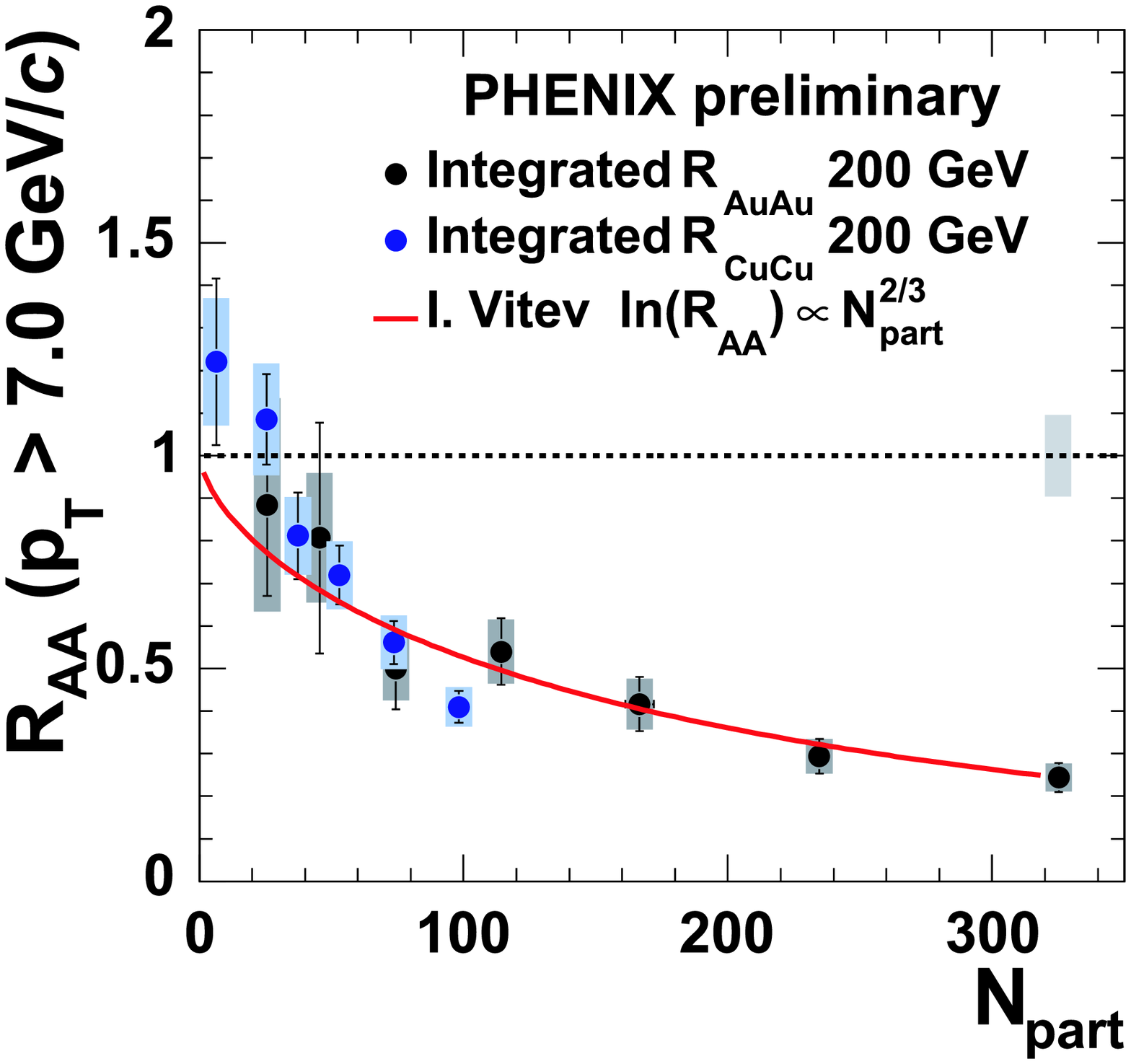}
   \end{center}
  \end{minipage}
  \begin{minipage}{0.45\linewidth}
   \begin{center}
    \includegraphics[height=.295\textheight]{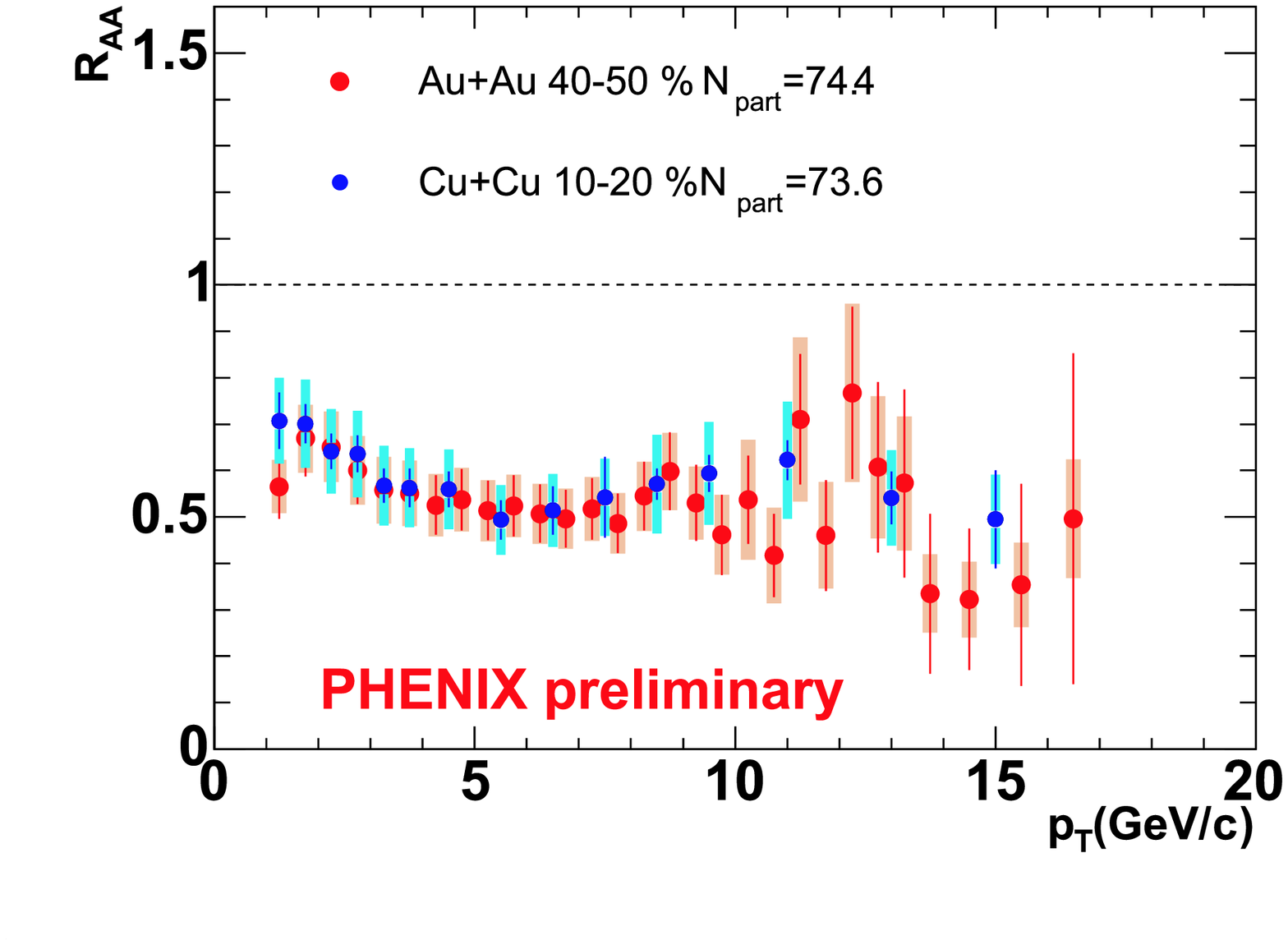}
   \end{center}
  \end{minipage}
 \caption{Left: Integrated R$_\mathrm{AA}$ as a function of the number
 of participant. Right: Comparison of R$_\mathrm{AA}$ between Au+Au and 
 Cu+Cu at the similar number of participant (N$_\mathrm{part} \sim$ 74).}
 \label{fig:comp}
\end{figure}

The PHENIX Run-5 Cu+Cu data may give us some insight into the relationship
between jet quenching effects and the properties of the medium through
the systematic study of high-p$_\mathrm{T}$ suppression in systems of
different size. 
Figure~\ref{fig:comp} shows the comparison of R$_\mathrm{AA}$ in Au+Au
collisions to that in Cu+Cu collisions.
As shown on the right panel of Fig.~\ref{fig:comp}, R$_\mathrm{AA}$ in Au+Au
collisions is very similar to that in Cu+Cu collisions for similar
number of participants.
Also, the integrated R$_\mathrm{AA}$ as a function of the number of 
participants (N$_\mathrm{part}$) in Au+Au collisions is very similar to
that in Cu+Cu collisions as shown in the left panel of
Fig.~\ref{fig:comp}. 
R$_\mathrm{AA}$ in the Au+Au and Cu+Cu systems are very similar and
follow the scaling with $\ln R_\mathrm{AA} \propto N_{part}^{2/3}$,
which is proposed based on a model where energy loss depends on  
the path-length of hard scattered partons going through the 1+1D
expanding medium~\cite{elosstest}.  

\section{Jet modification}

\begin{figure}[htb]
 \begin{center}
  \includegraphics[height=.31\textheight]{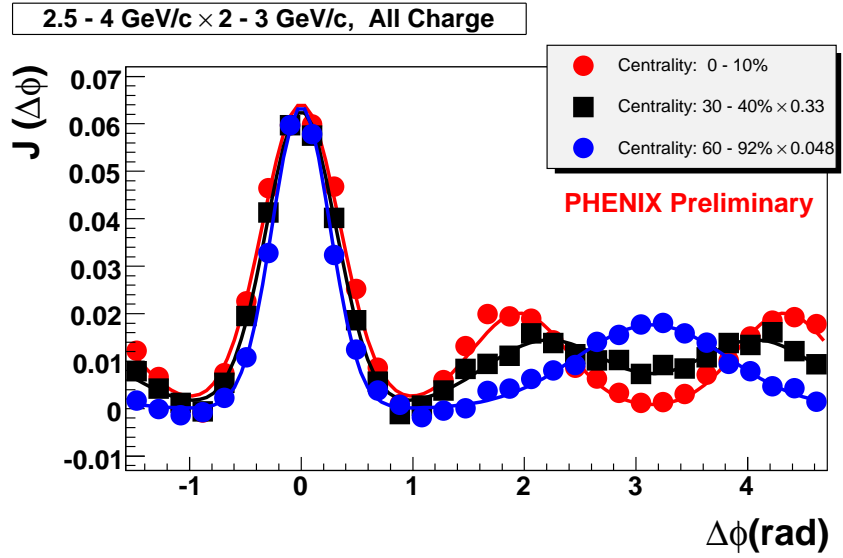}
  \caption{Correlation Function for central, mid-central and peripheral
  events after subtraction of the flow contribution.}
 \label{fig:jet_mod}
 \end{center}
\end{figure}

Two particle correlation is a powerful tool for understanding jets in
heavy ion collisions.    
Measurements of two-hadron azimuthal angular correlation functions at
the RHIC-STAR experiment show a clear reduction in strength of the di-jet
signal in central Au+Au collisions~\cite{star}.
Furthermore, the recent PHENIX result shows a ``dip'' type away side jet
shape. 
Figure~\ref{fig:jet_mod} shows a correlation function for charged
hadrons with a trigger particle in the intermediate p$_\mathrm{T}$ range
(2.5 $<$ p$_\mathrm{T} <$ 4.0~GeV/$c$) and the associated particle with
2.0 $<$ p$_\mathrm{T} <$ 3.0~GeV/$c$.
The elliptic flow contributions are estimated based on the measured
elliptic flow and subtracted. 
While a clear structure of a near and away side jet peaks can be seen in 
peripheral collisions, the away side peak becomes broader and develops a
``dip'' around $\delta\phi=\pi$. 
Since the hard scattered partons may propagate through the medium
radiating gluons and interacting with the created medium until they
fragment into jet clusters, the double peak structure in central
collisions suggests a jet modification induced by interaction between
the scattered partons and the medium.
There are several theoretical models to explain this ``dip'' type away
side peak.
In some of the models, partons with velocities larger than the speed of
sound in a QGP produce shock waves propagating in a Mach cone, or
produce gluon radiation in a Cherenkov cone, with respect to parton's
momentum~\cite{mach0,mach1,mach2,cherenkov}. 
The theoretical models for jet suppression via gluon radiation predict
an accompanying jet broadening. 

\section{Summary}
The PHENIX preliminary results on identified single high-p$_\mathrm{T}$
hadron production in Au+Au and Cu+Cu collisions and on particle azimuthal
correlations are shown. 

According to the GLV calculation, the suppression factor of neutral pion
supports the existence of bulk matter where initial gluon
density~(dN$^g$/dy) is more than 1100 in Au+Au collisions at
$\sqrt{s_\mathrm{NN}}$ = 200~GeV. 
If collisions with similar number of participants are compared, there is
little if any dependence of R$_\mathrm{AA}$ on system-size.

A clear ``dip'' type away side jet shape is observed with a trigger
particle in the intermediate pT range. It suggests the
modification of jets in dense matter.

\end{document}